\documentstyle[12pt]{article}
\begin{document}

\date{}

\everymath{\displaystyle}
\renewcommand{\theequation}{{\rm\thesection.\arabic{equation}}}

\begin{titlepage}
\title{THE PHENOMENA OF THE TIME-VIOLATING PHOTON POLARIZATION PLANE
AND NEUTRON SPIN ROTATION BY A DIFFRACTION GRATING.
NEW METHODS OF MEASURING OF THE TIME-VIOLATING INTERACTIONS.}
\author{
\bf V.G.BARYSHEVSKY\\ [0.6em]
\sl Nuclear Problems Institute, Bobruiskaya Str 11, \\
\sl Minsk 220080 Belarus. \\
\sl Electronic address: bar@inp.belpak.minsk.by \\
}
\maketitle

\begin{abstract}

It is shown, that the new phenomena of the T-violating
photon polarization plane and neutron spin rotation appear
under diffraction in a
noncentrosymmetrical diffraction grating. The equations describing the
T-violating photon scattering by a diffraction grating
and neutron interaction with noncentrosymmetrical crystal have been
obtained. These phenomena may be used for time-violating
interactions investigation.

\end{abstract} \thispagestyle{empty} \end{titlepage}
%-------------------------------------------------------------------

%\pagestyle{empty}
\centerline{\bf
THE PHENOMENA OF THE TIME-VIOLATING PHOTON POLARIZATION PLANE} %
\centerline{\bf
AND NEUTRON SPIN ROTATION BY A DIFFRACTION GRATING.} %
\centerline{\bf
NEW METHODS OF MEASURING OF THE TIME-VIOLATING INTERACTIONS.}\vspace{0.7cm} %
\centerline{V.G.BARYSHEVSKY } \centerline{Nuclear Problems Institute,
Bobruiskaya Str 11,} \centerline{Minsk 220080 Belarus.} %
\centerline{Electronic address: bar@inp.belpak.minsk.by } %
\vspace{0.7cm}

Since the discovery of the CP-violation in decay of K-mesons (Christenson,
Cronin, Fitch and Turlay (1964)), a few attempts have
been undertaken to observe experimentally this phenomenon in different
processes . However, that experiments have not been successful. At the
present time, novel more precise experimental schemes are actively
discussed: observation of the atom (Lamoreaux (1989)) and neutron (Forte (1983),
Fedorov , Voronin and Lapin (1992)) electric dipole
moment; the T(time)-violating atom's (molecule's) spin rotation in a laser
wave and the T-violating refraction of a photon in a polarized atomic or
molecular gas (Baryshevsky (1993, 1994)).

\section {P and T-violating photon polarisation plane rotation.}
\setcounter {equation}{0}
In accordance with Baryshevsky (1993, 1994) the P(parity)- and T-violating
dielectric permittivity tensor $\varepsilon _{ik}$
(for case$\varepsilon _{ik}-\delta _{ik} \ll 1$ ) is given by

\begin{equation}
\varepsilon _{ik}=\delta _{ik}+\chi _{ik}=\delta _{ik}+\frac{4\pi \rho }{k^2}%
f_{ik}\left( 0\right)  ,
\end{equation}

where $\chi _{ik}$ is the polarizability tensor of the matter, $\rho $ is
the number of atoms (molecules) per cm$^3$ , $k$ - the photon wave number.
The quantity $f_{ik}\left( 0\right)$ is the tensor part of the zero-angle
amplitude of elastic coherent scattering of a photon by an atom (molecule)
$f\left( 0\right) =f_{ik}\left( 0\right) e^{^{\prime ^{*}}}e_k$. Here $\vec e$
and $\vec e^{^{\ \prime }}$ are the polarization vectors of initial and
scattered photons. Indices i=1,2,3 are refered to coordinates x, y, z,
respectively, repeated indices imply summations. At zero external electric
and magnetic fields, the amplitude $f_{ik}\left( 0\right)$ can be written as

\begin{eqnarray}
f_{ik}\left( 0\right) &=&f_{ik}^{ev}\left( 0\right) +f_{ik}^{P,T}\left(
0\right) =f_{ik}^{ev}\left( 0\right) +\frac{\omega ^2}{c^2}[i\beta
_s^P\varepsilon _{ikl}n_l-\beta _\upsilon ^P\left( \left\langle \vec
J\right\rangle \vec n\right) \delta _{ik}+ \\
&&+i\beta _t^P\varepsilon _{iml}\left\langle \hat Q_{mk}\right\rangle
n_l+\frac 12\beta _t^T\left( \left\langle \hat Q_{im}\right\rangle
\varepsilon _{mlk}n_l+\left\langle \hat Q_{km}\right\rangle \varepsilon
_{mli}n_l\right) , \nonumber
\end{eqnarray}

where $f_{ik}^{ev}$ is the P, T even (invariant) part of $f_{ik}\left(
0\right) $, $f_{ik}^{P,T}$ is the P,T-violating part of $f_{ik}\left(
0\right) $, $\beta _{s,v,t}^{P,T}$ is the scalar (vector, tensor) P,T
violating polarizability of an atom (molecule), $\varepsilon _{ikl}$ is the
total antisymmetrical unit tensor of the rank three, $\vec n=\frac{\vec k}k$
, $\left\langle \vec J\right\rangle =Sp \rho _J\vec J$ , $\rho _J$
is the spin density matrix of an atom (molecule) with the spin $\vec J$, %
$\left\langle Q_{im}\right\rangle =Sp \rho _J\hat Q_{im}$. The
second rank tensor $\hat Q_{im}$ is given by
		  \
\[
\hat Q_{im}=\left[ 2J\left( 2J-1\right) \right] ^{-1}\left\{ \hat J_i\hat
J_m+\hat J_m\hat J_i-\frac 23J\left( J+1\right) \delta _{im}\right\}
\]

In view of (1.1,1.2), the T-violating processes affect upon the dielectric
permittivity $\varepsilon _{ik}$ and, as a result, upon the refraction index
only in media with polarized atoms (molecules) of the spin equaled or larger
than 1. This part of $\varepsilon _{ik}$ is proportional to $\beta _t^T$ .
If the atoms' (molecules') spins are nonpolarized, only the P-violating term
$f_{ik}\left( 0\right) =\frac{\omega ^2}{c^2}i\beta _s^P\varepsilon
_{ikl}n_l $ exists. The term proportional to $\beta _s^P$ describes the
P-violating rotation of a light polarization plane in metallic vapours
(Barkov and Zolotariov 1978, Bouchiat and Pottier 1986, Khriplovich 1991).

As it has been shown in (Baryshevsky 1993, 1994), when an atom interacts
with two coherent electromagnetic waves, the energy of this interaction
depends on the T-violating scalar polarizability $\beta _t^T$. Interaction
of an atom (molecule) with two waves can be considered as a process of
rescattering of one wave into another and vice versa. Then, as it follows
from an expression for the effective interaction energy, the amplitude $%
f\left( \vec k^{^{\prime }},\vec k\right) $ of the photon scattering by an
unpolarized atom (molecule) at a non-zero angle is given by (Baryshevsky
1994):

\begin{equation}
f\left( \vec k^{^{\prime }},\vec k\right) =f_{ik}e^{^{\prime ^{*}}}_i e_k=\frac{%
\omega ^2}{c^2}\left( \alpha _s\vec e^{^{\prime ^{*}}}\vec e+i\frac 12\beta
_s^P\left( \vec n^{^{\prime }}+\vec n\right) \left[ \vec e^{^{\prime
^{*}}}\vec e\right] +\beta _s^T\left( \vec n^{^{\prime }}-\vec n\right)
\left[ \vec e^{^{\prime ^{*}}}\vec e\right] \right) ,
\end{equation}

where $\vec k$ is the wave vector of a scattered photon, $\vec n^{^{\prime
}}=\frac{\vec k^{^{\prime }}}k$ , $\alpha _s$ is the scalar P,T-invariant
polarizability of an atom (molecule). Expression (1.3) holds true in the
absence of .external electric and magnetic fields.

It should be emphasized that expression (1.3) for the elastic scattering
amplitude can be derived from the general principles of symmetry. Indeed,
there are four independent unit vectors: $\vec \nu _1=\frac{\vec k^{^{\prime
}}+\vec k}{\left| \vec k^{^{\prime }}+\vec k\right| }$ , $\vec \nu _2=\frac{%
\vec k^{^{\prime }}-\vec k}{\left| \vec k^{^{\prime }}-\vec k\right| }$ , $%
\vec e$ and $\vec e^{^{\ \prime }}$ , which completely describe geometry of
the elastic scattering process. The elastic scattering amplitude $f\left(
\vec k^{^{\prime }},\vec k\right) $ depends on these vectors and therewith
is a scalar. Obviously, one can compose three independent scalars from these
vectors: $\vec e^{^{\ \prime }}\vec e$ , $\vec \nu _1\left[ \vec e^{^{\ \prime
^{*}}}\vec e\right] $ , $\vec \nu _2\left[ \vec e^{^{\ \prime ^{*}}}\vec
e\right] $ . As a result, the scattering amplitude can be written as:

\begin{equation}
f\left( \vec k^{^{\prime }},\vec k\right) =f_s\left( \vec k^{^{\prime
}},\vec k\right) \vec e^{^{\ \prime ^{*}}}\vec e+if_s^P\left( \vec k^{^{\prime
}},\vec k\right) \vec \nu _1\left[ \vec e^{^{\ \prime ^{*}}}\vec e\right]
+f_s^T\left( \vec k^{^{\prime }},\vec k\right) \vec \nu _2\left[ \vec
e^{^{\ \prime ^{*}}}\vec e\right] ,
\end{equation}

where $f_s$ is the P-,T- invariant scalar amplitude, $f_s^P$ is the
P-violating scalar amplitude, and $f_s^T$ is the P-,T- violating scalar
amplitude.

It can easily be found from (1.3,1.4) that the term proportional to $\beta
_s^T\left( f_s^T\right) $ vanishes in the case of forward scattering $\left(
\vec n^{^{\prime }}\rightarrow \vec n\right) $. Vice versa, in the case of
back scattering $\left( \vec n^{^{\prime }}\rightarrow -\vec n\right) $ the
term proportional to $\beta _s^P\left( f_s^P\right) $ gets equal to zero.

Thus, one can conclude that the T-violating interactions manifest themselves
in the processes of scattering by atoms (molecules). However, the scattering
processes are usually incoherent and their cross sections are too small to
hope for observation of the T-violating effect. Another situation takes
place for diffraction gratings in the vicinity of the Bragg resonance where
the scattering process is coherent. As a result, the intensities of
scattered waves strongly increase: for instance, in the Bragg (reflection)
diffraction geometry the amplitude of the diffracted-reflected wave may
reach the unity. It gives us an opportunity to study the T-violating
scattering processes (Baryshevsky 1994).

In the present paper, equations describing the T-violating scattering by a
diffraction grating have been obtained. It has been shown that the photon's
refraction index in a non-center-symmetrical grating depends on the
T-violating amplitude $f_s^T$ . It can result in a new phenomenon: the
T-violating rotation of the photon polarization plane. It has also been
shown that the rotation angle rises sharply in the back-scattering
diffraction geometry when the conditions of the photon resonance
transmission are satisfied.

\subsection *{The P-,T-violating electromagnetic waves diffraction by a
diffraction grating}

The phenomenon of P-, T- invariant diffraction of electromagnetic waves by
diffraction gratings has been studied in detail for a very wide range of
wavelengths (see, for examples, Shih-Lin Chang (1984), Tamir (1988),
Maksimenko and Slepyan (1997)). Accoding to these articles equations of the dynamic
diffraction can be derived from the Maxwell's equations if the
permittivity tensor $\varepsilon _{ik}\left( \vec r,\omega \right) $ of a
spatially periodic grating is known.

To include the P, T violating processes into the diffraction theory, let us
consider the microscopic Maxwell equations:

\begin{eqnarray}
curl \vec E &=&-\frac 1c\frac{\partial \vec B}{\partial t}\quad  ,  \quad
curl \vec B=\frac 1c\frac{\partial \vec E}{\partial t}+\frac{4\pi }%
c\vec j  \quad ,  \\
div \vec E &=&4\pi \rho \quad  , \quad  div \vec B=0 \quad , \quad \frac{%
\partial \rho }{\partial t}+ div \vec j=0  \quad .\nonumber
\end{eqnarray}

where $\vec E$ is the electric field strength and $\vec B$ is the magnetic field
induction, $\rho $ and $\vec j$ are the microscopic densities of the
electrical charge and the current induced by an electromagnetic wave, $c$ is
the speed of light. The Fourier transformation of these equations (i.e. $%
\vec E\left( \vec r,t\right) =\frac 1{2\pi ^4}\int \vec E\left( \vec
k,\omega \right) e^{i\vec k\vec r} e^{-i\omega t} d^3kd\omega $ and so on) leads us to the
equation for $\vec E\left( \vec k,\omega \right) $ as follows:

\begin{equation}
\left( -k^2+\frac{\omega ^2}{c^2}\right) \vec E\left( \vec k,\omega \right)
=-\frac{4\pi i\omega }{c^2}\left[ \vec j\left( \vec k,\omega \right) -\frac{%
c^2k^2}{\omega ^2}\vec n\left( \vec n\vec j\left( \vec k,\omega \right)
\right) \right] ,
\end{equation}

where $\vec n=\frac{\vec k}{k}$ .

In linear approximation, the current $\vec j\left( \vec r,\omega \right) $
is coupled with $\vec E \left( \vec r,\omega \right)$ by the well-known dependence:

\begin{equation}
j_i\left( \vec r,\omega \right) =\int d^3r^{^{\prime }}\sigma _{ij}\left(
\vec r,\vec r^{^{\prime }},\omega \right) E_j\left( \vec r^{^{\prime
}},\omega \right)
\end{equation}

with $\sigma _{ij}\left( \vec r,\vec r^{^{\prime }},\omega \right) $ as the
microscopic conductivity tensor being a sum of the conductivity tensors of
the atoms (molecules) constituting the diffraction grating:

\begin{equation}
\sigma _{ij}\left( \vec r,\vec r^{^{\prime }},\omega \right)
=\sum_{A=1}^N\sigma _{ij}^A\left( \vec r,\vec r^{^{\prime }},\omega \right)
\end{equation}

Here $\sigma _{ij}^A$ is the conductivity tensor of the A-type scatterers.
The summation is over all atoms (molecules) of the grating.

In a diffraction grating, the tensor $\sigma _{ij}\left( \vec r,\vec
r^{^{\prime }},\omega \right) $ is a spatially periodic function. It allows
one to derive the expansion of $j_i\left( \vec k,\omega \right) $ from (1.7)
as follows:

\begin{equation}
j_i\left( \vec k,\omega \right) =\frac 1{V_0}\sum_{\vec \tau }\sigma
_{ij}^c\left( \vec k,\vec k-\vec \tau ,\omega \right) E_j\left( \vec k-\vec
\tau ,\omega \right)
\end{equation}

where $\sigma _{ij}^c$ is the Fourier transform of the conductivity tensor
of a grating's elementary cell, $\vec \tau $ is the reciprocal lattice
vector of the diffraction grating. Using current representation (1.9), one can
obtain a set of equations from (1.6):

\begin{equation}
\left( -k^2+k_0^2\right) E_i\left( \vec k,\omega \right) =-\frac{\omega ^2}{%
c^2}\sum_{\vec \tau }\hat \chi _{ij}\left( \vec k,\vec k-\vec \tau \right)
E_j\left( \vec k-\vec \tau \right)
\end{equation}

Tensor of the diffraction grating susceptibility is given by

\begin{equation}
\hat \chi _{ij}\left( \vec k,\vec k-\vec \tau \right) =\left( \delta
_{il}-n_in_l\right) \chi _{lj}\left( \vec k,\vec k-\vec \tau \right)
\end{equation}

with

\[
\chi _{lj}\left( \vec k,\vec k-\vec \tau \right) =\frac{4\pi i}{V_0\omega }%
\sigma _{lj}\left( \vec k,\vec k-\vec \tau \right) =\frac{4\pi c^2}{%
V_0\omega ^2}F_{lj}\left( \vec k,\vec k-\vec \tau \right) .
\]

Here $F_{lj}\left( \vec k,\vec k-\vec \tau \right)=\frac{i \omega}{%
c^2}\sigma _{lj}\left( \vec k,\vec k-\vec \tau \right) $ is the amplitude of
coherent elastic scattering of an electromagnetic wave by a grating
elementary cell from a state with the wave vector $\vec k-\vec \tau $ to a
state with the wave vector $\vec k$.

The amplitude $F_{lj}$ is obtained by summation of atomic (molecular)
coherent elastic scatterig amplitudes over a grating's elementary cell:

\begin{equation}
F_{lj}\left( \vec k^{^{\prime }}=\vec k+\vec \tau ,\vec k\right)
=\left\langle \sum_{A=1}^{N_c}f_{lj}^A\left( \vec k^{^{\prime }}=\vec k+\vec
\tau ,\vec k\right) e^{-i\vec \tau \vec R_A}\right\rangle ,
\end{equation}

where $f_{lj}^A$ is the coherent elastic scattering amplitude by an A-type
atom (molecule), $\vec R_A$ is the gravity center coordinate of the A-type
atom (molecule) , $N_c$ is the number of the atoms (molecules) in an
elementary cell, angular brackets denote averaging over the coordinate
distribution of scatterers in a grating's elementary cell.

The amplitude $f_{lj}$ has been given by equation (1.4,1.3).

From (1.11), (1.12) and (1.4) one can obtaine an expression for the susceptibility
$\chi _{lj}$ of the elementary cell of an optically isotropic material:

\begin{equation}
\chi _{lj}\left( \vec k,\vec k-\vec \tau \right) =\chi _{s\vec \tau }\delta
_{lj}+i\chi _{s\vec \tau }^P\varepsilon _{ljf}\nu _{1f}^{\vec \tau }+\chi
_{s\vec \tau }^T\varepsilon _{ljf}\nu _{2f}^{\vec \tau }
\end{equation}

where

\[
\chi _{s\vec \tau }^{(P,T)}=\frac{4\pi c^2}{V_0\omega ^2}\left\langle
\sum_{A=1}^{N_c}f_{s}^{A(P,T)}\left( \vec k,\vec k-\vec \tau \right)
e^{-i\vec \tau \vec R_A}\right\rangle
\]

$\chi _{s\vec \tau }$ is the scalar P-, T- invariant susceptibility of an
elementary cell, $\chi _{s\vec \tau }^P$ is the P-violating, T- invariant
susceptibility of the elementary cell, and $\chi _{s\vec \tau }^T$ is the P-
and T- violating susceptibility of the elementary cell,

\[
\vec \nu _1^{\vec \tau }=\frac{2\vec k-\vec \tau }{\left| 2\vec
k -\vec \tau \right| }\quad , \quad \vec \nu _2^{\vec \tau }=\frac{\vec
\tau }\tau
\]

Then, using (1.10, 1.11,1.13) we can derive a set of equations describing the P
and T violating interaction of an electromagnetic wave with a diffraction
grating

\begin{eqnarray}
\left( -\frac{k^2}{k_0^2}+1\right) E_i\left( \vec k\right) &=&-\left( \delta
_{ij}-n_in_j\right) \chi _{_{s0}}E_j\left( \vec k\right) -i\chi _{s0}^P\left(
\delta _{il}-n_in_l\right) \varepsilon _{ljf}n_fE_j\left( \vec k\right) - \nonumber \\
&&-\sum_{\vec \tau \neq 0}\{\left( \delta _{ij}-n_in_j\right) \chi _{_{s\vec
\tau }} E_j\left( \vec k-\vec \tau \right) + \nonumber  \\
&&+i\chi _{_{s\vec \tau}}^P \left( \delta _{il}-n_in_l\right)
\varepsilon _{ljf}\nu _{1f}^{\vec \tau }E_j\left( \vec k-\vec \tau \right) +
\\
&&+\chi _{s_{\vec\tau}}^T\left( \delta _{il}-n_in_l\right) \varepsilon _{ljf}\nu
_{2f}^{\vec \tau }E_j\left( \vec k-\vec \tau \right) \}  , \nonumber
\end{eqnarray}

where $k_0=\frac \omega c$

Assuming the interaction to be P, T invariant $\left( \chi _s^P=\chi
_s^T=0\right) $, equations (1.14) reduce to the conventional set of equations
of dynamic diffraction theory (Shih-Lin Chang (1984)).

\subsection *{The phenomenon of T-violating rotation of the photon polarization
plane by a diffraction grating}

Let us suppose, first of all, the photon $\omega $ frequency and the wave
vector $\vec k$ to be such that the Bragg diffraction conditions $\vec
k=\vec k\pm \vec \tau $ and $\left| \vec k^{^{\prime }}\right| =\left| \vec
k\right| $ are not fulfilled exactly, and the inequality $\frac{k_0^2\left|
\hat \chi _{lj}\left( \vec k,\vec k-\vec \tau \right) \right| }{\left( \vec
k-\vec \tau \right) ^2-k_0^2}\ll 1$ holds true. In this case, the diffracted
wave amplitude is much less comparing with the transmitted one: $\left| \vec
E\left( \vec k-\vec \tau \right) \right| \ll \left| \vec E\left( \vec
k\right) \right| $ , and the perturbation theory can be applied for the
further analysis. As a result in the first approximation of the perturbation
theory one can derive from (1.10) that

\begin{equation}
E_j\left( \vec k-\vec \tau \right) \simeq \frac{k_0^2}{\left( \vec k-\vec
\tau \right) ^2-k_0^2}\hat \chi _{jf}\left( \vec k-\vec \tau ,\vec k\right)
E_f\left( \vec k\right) =\alpha _\tau ^{-1}\hat \chi _{jf}\left( \vec k-\vec
\tau ,\vec k\right) E_f\left( \vec k\right) ,
\end{equation}

where $\alpha _\tau =\frac{\vec \tau \left( \vec \tau -2\vec k_0\right) }{%
k_0^2}$ , $\vec k_0=\frac \omega c\vec n$ , $\left| \hat \chi _{ij}\right|
\ll 1$.

Substitution (1.15) into (1.10) results in the diffraction equations as follows

\begin{equation}
\left( k^2-k_0^2\right) E_i\left( \vec k,\omega \right) -k_0^2[\hat \chi
_{_{if}} \left( \vec k,\vec k\right) +\sum_{\vec \tau \neq 0}\alpha _\tau
^{-1}\hat \chi _{_{ij}}\left( \vec k,\vec k-\vec \tau \right) \hat \chi
_{_{jf}}\left( \vec k-\vec \tau ,\vec k\right) ]E_f(\vec k,\omega )=0
\end{equation}

which can be rewritten in more simple form

\begin{equation}
\left( k^2-k_0^2\hat \varepsilon _{_{if}}\left( \vec k,\omega \right) \right)
E_f\left( \vec k,\omega \right) =0
\end{equation}

by introducing the effective permittivity tensor

\begin{equation}
\hat \varepsilon _{if}\left( \vec k,\omega \right) =\delta _{if}+\hat \chi
_{_{jf}}\left( \vec k,\vec k,\omega \right) +\sum_{\vec \tau \neq 0}\alpha
_\tau ^{-1}\hat \chi _{_{ij}}\left( \vec k,\vec k-\vec \tau \right) \hat \chi
_{_{jf}}\left( \vec k-\vec \tau ,\vec k\right)
\end{equation}

One can see that even far away from the exact Bragg conditions, where the
diffracted wave amplitudes are small, a spatially periodic isotropic medium
manifests the optical anisotropy being characterized by the effective
permittivity tensor $\hat \varepsilon _{if}\left( \vec k,\omega \right) $.

Let a photon be incident on a grating normally to its reflection planes. In
other words, let the photon wave vector $\vec k_{0}$ be antiparallel
to the reciprocal lattice vector $\vec \tau $ , i.e. $\vec k_0\uparrow
\downarrow \vec \tau $ . In this case, the back-scattering diffraction
regime can be realized for photons with wave numbers defined by the relation
$k\approx \frac 12\tau $ . If, nevertheless, the inequality $\alpha _\tau
\gg \left| \chi _{ij}\left( \vec k,\vec k-\vec \tau \right) \right| $ holds
true, we can use set of equations (1.16) in which there is only one term
satisfying the conditions $\vec \tau \uparrow \downarrow \vec k$ and $\tau
\simeq 2k$ .

Let the coordinate axis $z$ be parallel to $\vec k_0$ , $\vec\tau $ . In this
case, the tensor $\hat \chi _{ij}$ has nonzero components at $i$ , $j$ = 1,2
only. As a result, set of equations (1.17) can be rewritten in the form as
follows:

\begin{equation}
\left( k^2-k_0^2\varepsilon _{ij}\left( \vec k,\omega \right) \right)
E_j\left( \vec k,\omega \right) =0
\end{equation}

with the permittivity tensor given by

\begin{eqnarray}
\varepsilon _{ij}\left( \vec k,\omega \right) &=&\varepsilon _0\delta
_{ij}+i\chi _s^P\left( 0\right) \varepsilon _{ij3}n_3+ \\
&&+\alpha _\tau ^{-1}\left[ \chi _s\left( \vec \tau \right) \chi _s^T\left(
-\vec \tau \right) -\chi _s\left( -\vec \tau \right) \chi _s^T\left( \vec
\tau \right) \right] \varepsilon _{ij3}\nu _{23}^{\vec \tau }  \nonumber
\end{eqnarray}

In the above equations we introduced the designations:

\begin{eqnarray*}
\chi \left( \vec \tau \right) &=&\chi \left( \vec k,\vec k-\vec \tau \right)
\quad , \quad \chi \left( -\vec \tau \right) =\chi \left( \vec k-\vec \tau ,\vec
k\right) \quad , \quad \varepsilon _0=1+\chi _s^{eff} \quad ,\\
\chi _s^{eff} &=&\chi _s\left( 0\right) -\alpha _\tau ^{-1}\chi _s\left(
\vec \tau \right) \chi _s\left( -\vec \tau \right) \quad , \quad \vec n=\frac{\vec k%
}k\quad , \quad \vec \nu _2^\tau =\frac{\vec \tau }\tau \quad .
\end{eqnarray*}

The term proportional $t_0$ $\chi _s^P\left( 0\right) $ describes the
P-violating and T-invariant rotation of the light polarization plane about
the direction $\vec n$. This term does not depend on a structure of the
diffraction grating and exists for any ordinary spatially isotropic media.
Unlike to that, the term proportional to $\chi _s^T$ is T-violating and
depends on the grating's structure. The term proportional $\chi _s^T$ looks
like the term proportional to $\chi _s^P$ and is responsible for the
polarization plane rotation about $\vec \nu _2^\tau $.

It is known that the phenomenon of the polarization plane rotation arises
when right- and left-circularly polarized photons have different indices of
refraction in a medium $n_{+}$ and $n_{-}$, respectively. It means that the
tensor $\varepsilon _{ij}$ is diagonal for a given circular polarization
and, concequently, the set of equations (1.19) is split into two independent
equations. Really, let us write (1.19) in the vector notation:

\begin{equation}
\left( k^2-k_0^2\varepsilon _0\right) \vec E-ik_0^2\chi _s^P\left( 0\right)
\left[ \vec E\vec n\right] -k_0^2\alpha _\tau ^{-1}\left[ \chi _s\left( \vec
\tau \right) \chi _s^T\left( -\vec \tau \right) -\chi _s\left( -\vec \tau
\right) \chi _s^T\left( \vec \tau \right) \right] \left[ \vec E\vec \nu
_2^\tau \right]
\end{equation}

and let $\vec e_1$ be the unit polarization vector of a linearly polarized
photon; $\vec e_2=\left[ \vec n\vec e_1\right] $ , $\vec e_1\perp \vec
e_2\perp \vec n$ . Then, the unit vectors corresponding to the circular
polarizations are as follows $\vec e_{\pm}=\frac{\vec e_1 \pm i\vec e_2}{\sqrt{2}}$%
. For the right $\left( \vec e_{+}\right) $ , left $\left( \vec e_{-}\right) $
circularly polarized photons, the field $\vec E$ can be represented by $\vec
E=c_{\left( \pm \right) }\vec e_{\left( \pm \right) }$. As a result, it follows
from (1.19, 1.21) that:

\begin{equation}
\left( k^2-k_0^2\varepsilon _0\right) c_{\pm }\pm k_0^2\chi _s^P\left(
0\right) c_{\pm }\pm ik_0^2\alpha _\tau ^{-1}\left[ \chi _s\left( \vec \tau
\right) \chi _s^T\left( -\vec \tau \right) -\chi _s\left( -\vec \tau \right)
\chi _s^T\left( \vec \tau \right) \right] c_{\pm }=0
\end{equation}

The corresponding refractive indices are obtained from (1.20)

\begin{equation}
n_{\pm }^2=\frac{k^2}{k_0^2}=\varepsilon _0\mp \chi _s^P\left( 0\right) \mp
i\alpha _\tau ^{-1}\left[ \chi _s\left( \vec \tau \right) \chi _s^T\left(
-\vec \tau \right) -\chi _s\left( -\vec \tau \right) \chi _s^T\left( \vec
\tau \right) \right]
\end{equation}

The angle of the photon polarization plane rotation is defined by :

\begin{equation}
\vartheta =k_0 Re \left( n_{+}-n_{-}\right) L_{ph}
\end{equation}

where $L_{ph}$ is the photon propagation length in the medium, $Re n_{\pm}$ is the
real part of $n_{\pm }$. Then, the expression for $\vartheta $ can easily be
derived from (1.24,1.23)

\begin{eqnarray}
\vartheta &=&k_0 Re \frac{n_{+}^2-n_{-}^2}{n_{+}+n_{-}}L\simeq -k_0%
Re \chi _s^P\left( 0\right) L- \\
&&-k_0\ Re i\alpha _\tau ^{-1}\left[ \chi _s\left( \vec \tau \right)
\chi _s^T\left( -\vec \tau \right) -\chi _s\left( -\vec \tau \right) \chi
_s^T\left( \vec \tau \right) \right] L  \nonumber
\end{eqnarray}

One can conclude, thus, that the T-violating interaction results in the
phenomenon of the T-violating rotation of the photon polarization plane. The
effect manifests itself when the condition $Re i\left[ \chi
_s\left( \vec \tau \right) \chi _s^T\left( -\vec \tau \right) -\chi _s\left(
-\vec \tau \right) \chi _s^T\left( \vec \tau \right) \right] \neq 0$ holds
true.

It follows from (1.13) that the susceptibilities $\chi _s^{P,T}\left( \vec
\tau \right) $ can be presented as

\begin{eqnarray}
\chi _s^{P,T}\left( \vec \tau \right) &=&\chi _{1s}^{P,T}\left( \vec \tau
\right) -\chi _{2s}^{P,T}\left( \vec \tau \right) =\left\langle
\sum_{A=1}^{N_c}f_{s}^{A(P,T)}\left( \vec k,\vec k-\vec \tau \right) \cos
\vec \tau \vec R_A\right\rangle - \\
&&-i\left\langle \sum_{A=1}^{N_c}f_{s}^{A(P,T)}\left( \vec k,\vec k-\vec
\tau \right) \sin \vec \tau \vec R_A\right\rangle  \nonumber
\end{eqnarray}

where

\begin{equation}
\chi _{1s}^{P,T}\left( \vec \tau \right) =\chi _{1s}^{P,T}\left( -\vec \tau
\right) \qquad \chi _{2s}^{P,T}\left( \vec \tau \right) =-\chi
_{2s}^{P,T}\left( -\vec \tau \right)
\end{equation}

In view of (1.26,1.27) we can rewrite (1.25) as:

\begin{equation}
\vartheta =-k_0 Re \chi _s^P\left( 0\right) L+2k_0\alpha _\tau ^{-1}%
 Re \left[ \chi _{1s}\left( \vec \tau \right) \chi _{2s}^T\left(
\vec \tau \right) -\chi _{2s}\left( \vec \tau \right) \chi _{1s}^T\left(
\vec \tau \right) \right] L
\end{equation}

So, the T-violating rotation arises in the case of nonzero odd part of the
suscetibility: $\chi _2\left( \vec \tau \right) \neq 0$. Such a situation is
possible if an elementary cell of the diffraction grating does not posses
the center of symmetry.

In accordance with (1.28), the angle of the T-violating rotation grows at $%
\alpha _\tau \rightarrow 0$. .However, the condition $\alpha _\tau
\left| \chi _s\left( \vec \tau \right) \right| \ll 1$ violates at $%
\alpha _\tau ^{-1}\rightarrow 0$, where the amplitude of diffracted and
transmitted waves are comparable: $E\left( \vec k-\vec \tau \right) \simeq
E\left( \vec k\right) $ and, consequently, the perturbation theory gets
unapplicable. A rigorous dynamical diffraction theory must be applied in
this case.

\subsection *{The T-violating polarization plane rotation in the Bragg
diffraction scheme}

Let the Bragg condition is fulfilled for the only diffracted wave and is
violated for all other possible ones. It allows us to restrict ourselves to
the two-wave approximation of the dynamical diffraction theory (Shih-Lin
Chang (1984)). In that case, set of equations (1.14) reduces to two coupled
equations, which for the back-scattering diffraction scheme $\left( \vec
k_0\parallel \vec \tau \right) $ take the form as follows:

\begin{eqnarray*}
\left( \frac{k^2}{k_0^2}-1\right) E_j\left( \vec k\right) &=&\chi _s\left(
0\right) E_j\left( \vec k\right) +i\chi _s^P\left( 0\right) \varepsilon
_{jmf}E_m\left( \vec k\right) n_f+ \\
&&+\chi _s\left( \vec \tau \right) E_j\left( \vec k-\vec \tau \right) +\chi
_s^T\left( \vec \tau \right) \varepsilon _{jmf}E_m\left( \vec k-\vec \tau
\right) \nu _{2f}^{\vec \tau } ,
\end{eqnarray*}
\hspace{14cm}
(1.29)
\begin{eqnarray*}
\left( \frac{\left( \vec k-\vec \tau \right) ^2}{k_0^2}-1\right) E_j\left(
\vec k-\vec \tau \right) &=&\chi _s\left( 0\right) E_j\left( \vec k-\vec
\tau \right) + \nonumber\\
&&i\chi _s^P\left( 0\right) \varepsilon _{jmf}n_f\left( \vec
k-\vec \tau \right) E_m \left( \vec k-\vec \tau \right) + \nonumber\\
&&+\chi _s\left( -\vec \tau \right) E_j\left( \vec k\right) +\chi _s^T\left(
-\vec \tau \right) \varepsilon _{jmf}\nu _{2f}^{-\vec \tau }E_m\left( \vec
k\right) ,
\end{eqnarray*}

\addtocounter{equation}{1}

$\vec n\left( \vec k-\vec \tau \right) =\frac{\vec k-\vec \tau }{\left| \vec
k-\vec \tau \right| }$ .

Based on the above consideration, we can conclude that set of equations (1.29)
can be diogonalized for the photon of a given circular polarization. Let the
right-circularly polarized photon $\left( \vec e_{+}\right) $ be incident on
the diffraction grating. The diffraction process, as it follows from (29),
results in the appearance of a back-scattered photon with the left circular
polarization $\left( \vec e_{-}^{\vec \tau }\right) $. This is because the
momentum of the back-scattered photon $\vec k^{^{\prime }}=\vec k-\vec \tau $
is antiparallel to the momentum $\vec k$ of the incident one. It is obvious
that the left-circularly polarized photon will produce a right-circularly
polarized back-scattered one.

Thus, for circularly polarized photons set of vector equations (1.29) can be
split into two independent sets of scalar equations:

\begin{eqnarray}
\left( \frac{k^2}{k_0^2}-1\right) C_{\pm }\left( \vec k\right) &=&\left(
\chi _s\left( 0\right) \mp \chi _s^P\left( 0\right) \right) C_{\pm }\left(
\vec k\right)+ \nonumber \\
&&+\left( \chi _s\left( \tau \right) \mp i\chi _s^T\left( \tau \right) \right)
C_{\mp }\left( \vec k-\vec \tau \right)   \\
\left( \frac{\left( \vec k-\vec \tau \right) }{k_0^2}-1\right) C_{\mp
}\left( \vec k-\vec \tau \right) &=&\left( \chi _s\left( 0\right) \pm \chi
_s^P\left( 0\right) \right) C_{\mp }\left( \vec k-\vec \tau \right) +
\nonumber \\
&&+\left( \chi _s\left( -\vec \tau \right) \pm i\chi _s^T\left( -\vec \tau
\right) \right) C_{\pm }\left( \vec k\right)  \nonumber
\end{eqnarray}

Note that equations (1.30) are identical in form to conventional equations of
the two-wave dynamical diffraction (Shih-Lin Chang (1984)):

\begin{eqnarray}
\left( \frac{k^2}{k_0^2}-1\right) c_0 &=&\chi _0c_0+\chi _\tau c_\tau \\
\left( \frac{k_\tau ^2}{k_0^2}-1\right) c_\tau &=&\chi _0c_\tau +\chi
_{-\tau }c_0  \nonumber
\end{eqnarray}

It allows us to write down a solution immediately, without deriving (see,
for example, (Shih-Lin Chang (1984))). As a result, the amplitude of the
transmitted electromagnetic wave at the output is given by

\begin{equation}
\vec E_{\pm }=\vec e_{\pm }C_{\pm }\left( L\right) e^{i\vec k_0\vec r} ,
\end{equation}

\begin{eqnarray}
C_{\pm }\left( L\right) &=&2\left( \varepsilon _1^{\pm }-\varepsilon _2^{\pm
}\right) e^{i\frac 12k_0\varepsilon ^{\pm }L}[\left( 2\varepsilon _1^{\pm
}-\chi _0^{\pm }\right) e^{i\frac 12k_0\left( \varepsilon _1^{\pm
}-\varepsilon _2^{\pm }\right) L} - \\
&&-\left( 2\varepsilon _2^{\pm }-\chi _0^{\pm }\right) e^{-i\frac 12\left(
\varepsilon _1^{\pm }-\varepsilon _2^{\pm }\right) L}]^{-1} , \nonumber
\end{eqnarray}

where L is the thickness of the diffraction grating,

\[
\chi _0^{\pm }=\chi _s\left( 0\right) \mp \chi _s^P\left( 0\right) ,
\]

\begin{eqnarray}
\varepsilon _1^{\pm } &=&\frac 14\left[ \mp 2\chi _s^P\left( 0\right)
+\alpha \right] +\frac 14\{\left( 2\chi _s\left( 0\right) -\alpha \right)
^2-4\left( \chi _{1s}^2\left( \vec \tau \right) +\chi _{2s}^2\left( \vec
\tau \right) \right) \pm \nonumber\\
&&\pm 8\left[ \chi _{1s}\left( \vec \tau \right) \chi _{2s}^T\left( \vec
\tau \right) -\chi _{2s}\left( \vec \tau \right) \chi _{1s}^T\left( \vec
\tau \right) \right] \}^{\frac 12} ,
\end{eqnarray}

\begin{eqnarray*}
\varepsilon _2^{\pm } &=&\frac 14\left[ \mp 2\chi _s^P\left( 0\right)
+\alpha \right] -\frac 14\{\left( 2\chi _s\left( 0\right) -\alpha \right)
^2-4\left( \chi _{1s}^2\left( \vec \tau \right) +\chi _{2s}^2\left( \vec
\tau \right) \right) \pm \\
&&\pm 8\left[ \chi _{1s}\left( \vec \tau \right) \chi _{2s}^T\left( \vec
\tau \right) -\chi _{2s}\left( \vec \tau \right) \chi _{1s}^T\left( \vec
\tau \right) \right] \}^{\frac 12} ,
\end{eqnarray*}

\begin{eqnarray}
\alpha &\equiv &\alpha _\tau , \\
\varepsilon ^{\pm } &=&\varepsilon _1^{\pm }+\varepsilon _2^{\pm }=\mp \chi
_s^P\left( 0\right) +\frac \alpha 2 \nonumber ,
\end{eqnarray}

\begin{eqnarray}
\varepsilon _1^{\pm }-\varepsilon _2^{\pm } &=&\frac 12\{\left( 2\chi
_s\left( 0\right) -\alpha \right) ^2-4\left( \chi _{1s}^2\left( \vec \tau
\right) +\chi _{2s}^2\left( \vec \tau \right) \right) \pm \\
&&\pm 8\left[ \chi _{1s}\left( \vec \tau \right) \chi _{2s}^T\left( \vec
\tau \right) -\chi _{2s}\left( \vec \tau \right) \chi _{1s}^T\left( \vec
\tau \right) \right] \}^{\frac 12} \nonumber .
\end{eqnarray}

Consider now the diffraction of a photon with linear polarization $\vec e_1$
being the superposition of two opposite circular polarizations . In this
case

\begin{equation}
\vec E\left( \vec r\right) =\vec e_1e^{i\vec k_0\vec r}=\frac{\vec
e_{+}+\vec e_{-}}{\sqrt{2}}e^{i\vec k_0\vec r}
\end{equation}

and the amplitude of the transmitted wave $\vec E^{^{\prime }}\left(
r\right) $ can be presented by the superposition

\begin{eqnarray}
\vec E^{^{\prime }}\left( \vec r\right) &=&\left( \frac 1{\sqrt{2}}\vec
e_{+}C_{+}\left( L\right) +\frac 1{\sqrt{2}}\vec e_{-}C_{-}\left( L\right)
\right) e^{i\vec k_0\vec r}=  \nonumber\\
&=&\left( \frac{\vec e_1+i\vec e_2}2C_{+}\left( L\right) +\frac{\vec
e_1-i\vec e_2}2C_{-}\left( L\right) \right) e^{i\vec k_0\vec r}=
\\
&=&\left( \frac{C_{+}+C_{-}}2\vec e_1+i\frac{C_{+}-C_{-}}2\vec e_2\right)
e^{i\vec k_0\vec r} . \nonumber
\end{eqnarray}

In the case under consideration $c_{+}\neq c_{-}$.As follows from (1.38), this
results in changing of the photon polarization at the output.

Let us analyze expression (1.32) for the transmitted wave amplitude more
attentively. According to (1.32), the amplitude oscillates as a function of $%
\alpha $, i.e., as a function of the wavelength, with maximums in points
defined by the condition $k_0 Re \left( \varepsilon _1^{\pm
}-\varepsilon _2^{\pm }\right) L=2\pi m$ or $ Re \left( \varepsilon
_1^{\pm }-\varepsilon _2^{\pm }\right) =\frac{2\pi m}{k_0L}$with $m$ as an
integer.

Note first of all that the condition $m=0$ dictates at $\varepsilon _1^{\pm
}-\varepsilon _2^{\pm }\rightarrow 0$ the limit transition

\begin{equation}
\left\{ \left( 2\chi _s\left( 0\right) -\alpha \right) ^2-4\left( \chi
_{1s}^2\left( \vec \tau \right) +\chi _{2s}^2\left( \vec \tau \right)
\right) \pm 8\left[ \chi _{1s}\left( \vec \tau \right) \chi _{2s}^T\left(
\vec \tau \right) -\chi _{2s}\left( \vec \tau \right) \chi _{1s}^T\left(
\vec \tau \right) \right] \right\} ^{\frac 12}\rightarrow 0
\end{equation}

which determines the thresholds of the total Bragg reflection band where
there is a quickly damped inhomogeneous wave inside the diffraction grating.
As a result, the transmitted wave amplitude is small.

Let now $m\neq 0$. As we know, the T-violating interactions are very small: $%
\chi _{s1,2}^T\ll \chi _{s1,2}$. It allows one to expand the square root
into the Teylor series truncated beyond the second term:

\begin{equation}
\varepsilon _1^{\pm }-\varepsilon _2^{\pm }\simeq \left( \varepsilon
_1-\varepsilon _2\right) _{ev}\pm \frac{\chi _{1s}\left( \vec \tau \right)
\chi _{2s}^T\left( \vec \tau \right) -\chi _{2s}\left( \vec \tau \right)
\chi _{1s}^T\left( \vec \tau \right) }{\left( \varepsilon _1-\varepsilon
_2\right) _{ev}}
\end{equation}

where $\left( \varepsilon _1-\varepsilon _2\right) _{ev}=\frac 12\left\{
\left( 2\chi _s\left( 0\right) -\alpha \right) ^2-4\left( \chi _{1s}^2\left(
\vec \tau \right) +\chi _{2s}^2\left( \vec \tau \right) \right) \right\}
^{\frac 12}$  .

Let $ Re \chi _{_{s1,2}}\gg  Im \chi _{s1,2}$, i.e., the
absorption is assumed to be sufficiently small to satisfy the condition $k_0%
\ Im \left( \varepsilon _1-\varepsilon _2\right) _{ev}L\ll 1$ which
admits the consideration of the diffraction grating as an optically
transparent medium with $\left( \varepsilon _1-\varepsilon _2\right) _{ev}$
and $\chi _s$ as real functions. Let now the condition $k_0\left(
\varepsilon _1-\varepsilon _2\right) _{ev}L=2\pi m$ be fulfilled at $m\neq 0$%
. This condition defines the resonance transmission in the grating and
allows us to rewrite formula (1.40) in the form as follows

\begin{equation}
\varepsilon _1^{\pm }-\varepsilon _2^{\pm }=\frac{2\pi m}{k_0L}\pm \frac{%
k_0\left[ \chi _{1s}\left( \vec \tau \right) \chi _{2s}^T\left( \vec \tau
\right) -\chi _{2s}\left( \vec \tau \right) \chi _{1s}^T\left( \vec \tau
\right) \right] L}{2\pi m}
\end{equation}

By substituting (1.41) into (1.33,1.32) one can express $\vec E_{\pm }$ by

\begin{equation}
\vec E_{\pm }=\vec e_{\pm }\left( -1\right) ^m\left( 1-i\frac{k_0\left(
\alpha _{1,2}-2\chi _s\left( 0\right) \right) L}{8\pi m}k_0\Delta ^{\pm
}L\right) e^{ik_0\frac 12\varepsilon ^{\pm }\left( \alpha _{1,2}\right) L}
\end{equation}

where

\[
\alpha _{1,2}=2\chi _s\left( 0\right) \pm \sqrt{4\left( \chi _{1s}^2+\chi
_{2s}^2\right) +\left( \frac{4\pi m}{k_0L}\right) ^2}
\]

\[
\varepsilon ^{\pm }\left( \alpha _{1,2}\right) =\mp \chi _s^P\left( 0\right)
+\frac 12\alpha _{1,2}
\]

In view of that the second term in (1.42) is small and expression (1.42) can be
written as

\begin{equation}
\vec E_{\pm }=\vec e_{\pm }\left( -1\right) ^me^{i\varphi _{\pm }}
\end{equation}

where the phase terms are given by

\begin{equation}
\varphi _{\pm }=k_0\left[ \frac 12\varepsilon ^{\pm }\left( \alpha
_{1,2}\right) -\frac{k_0\left( \alpha _{1,2}-2\chi _s\left( 0\right) \right)
L}{8\pi m}\Delta ^{\pm }\right] L
\end{equation}

Using this equation one can find the angle of the polarization plane
rotation:

\begin{equation}
\vartheta = Re \left( \varphi _{+}-\varphi _{-}\right) =\vartheta
^P+\vartheta _{1,2}^T
\end{equation}

where the first term in the right-hand part defines the P-violating
T-invariant rotation angle:

\begin{equation}
\vartheta ^P=-k_0 Re\chi _s^P\left( 0\right) L
\end{equation}

and the second one corresponds to the T-violating rotation:

\begin{eqnarray}
\vartheta _{1,2}^T\left( \alpha _{1,2}\right) &=&\mp \frac{k_0^3L^3}{8\pi
^2m^2}\sqrt{4\left( \chi _{1s}^2+\chi _{2s}^2\right) +\left( \frac{4\pi m}{%
k_0L}\right) ^2}\times \\
&&\times \left[ \chi _{1s}\left( \vec \tau \right) Re \chi
_{2s}^T\left( \vec \tau \right) -\chi _{2s}\left( \vec \tau \right)
Re \chi _{1s}^T\left( \vec \tau \right) \right] , \nonumber
\end{eqnarray}

the $ sign \left( -\right) $ is for $\alpha _1$ , the $ sign %
\left( +\right) $ is for $\alpha _2$ .

The imaginary part of the T-violating polarizability $ Im \chi
_{s1,2}^T$ is responsible for the T-violating circular dichroism. Due to
that process, a linearly polarized photon gets a circular polarization at
the diffraction grating's output. The degree of the circular polarization of
the photon is determined from the relation:

\begin{eqnarray}
\delta _{1,2} &=&\frac{\left| \vec E_{+}\right| ^2-\left| \vec E_{-}\right|
^2}{\left| \vec E_{+}\right| ^2+\left| \vec E_{-}\right| ^2}\simeq %
Im \varphi _{-}- Im \varphi _{+}=k_0 Im \chi _s^P\left(
0\right) L\pm \\
&&\pm \frac{k_0^3L^3}{8\pi ^2m^2}\sqrt{4\left( \chi _{1s}^2+\chi
_{2s}^2\right) +\left( \frac{4\pi m}{k_0L}\right) ^2}\left[ \chi _{1s}\left(
\vec \tau \right) Im \chi _{2s}^T\left( \vec \tau \right) -\chi
_{2s}\left( \vec \tau \right) Im \chi _{1s}^T\left( \vec \tau
\right) \right]  \nonumber
\end{eqnarray}

It should be pointed out that the resonance transmission condition is
satisfied at a given m for two different values of $\alpha $. This is
because there is a possibility to approach to the Brilluan (the total Bragg
reflection) bandgap both from high and low frequencies. The T-violating
parts of the rotation angle are opposite in sign for $\alpha _1$ and for $%
\alpha _2$. It gives the addition opportunity to distinguish the T-violating
rotation from the P-violating T-invariant rotation. Indeed, the P-violating
rotation does not depend on the back Bragg diffraction in the general case
because the P-violating scattering amplitude equals zero for back scattering
(see (1.32-1.35).

In accordance with (1.47,1.48) the T-violating rotation and dichroism grow
sharply in the vicinity of the resonance Bragg transmission. At the first
glance, one could expect for $\vartheta ^T$ the dependence $\vartheta ^T\sim
k_0 Re \chi _{s1,2}^T\left( \vec \tau \right) L$ (see (1.25)).
However, in the vicinity of resonance, the rotation angle $\vartheta ^T$
turns out to be multiplied by the factor
$A=(8 \pi ^2 m^2)^{-1}k_0\sqrt{4\left( \chi
_{1s}^2+\chi _{_{2s}}^2\right) +\left( \frac{4\pi m}{k_0L}\right) ^2}Lk_0\chi
_{s1,2}L$ which provides the above mentioned growth (for example, $%
A\sim 10^5$ at $\chi _s\simeq 10^{-1}$ , $k_0\simeq 10^4\div 10^5cm^{-1}$ , $%
L=1cm$ , $m=1$ ).

There are different types of diffraction gratings destined for use in
optical and more longwave ranges. However, it should be noted that the
successful observation of the P-violating rotation has been performed by
means of studying of light transmission through gas targets . There are a
lot of theoretical calculations for atoms of such gases: see, for example,
(Khriplovich (1991)) for $Bi$ , $Tl$ , $Pb$ , $Dy$. From that point of view,
it would be preferable to use gases for studying of the T-violating
phenomena of polarization plane rotation and dichroism applying the
experience accumulated earlier. At the first glance, there is a serious
problem how to create such a diffraction grating in a gas. Nevertheless, the
problem can be solved if we make use some well-known results of the
electromagnetic theory of waveguides (Jackson (1962), Tamir (1988),
Maksimenko and Slepyan (1997)). According to this theory, there is a
correspondence between wave processes in waveguides with periodically
modulated boundaries and homogeneous filling and such processes in regular
waveguides filled by a periodic and, generally, anisotropic medium . Let us
consider a regular waveguide constituted by two plane-parallel surfaces; for
example, two metallic or dielectric mirrors. Let us then place a plane
diffraction grating on the surface of the mirror  and
fill the waveguide by the studied gas. Because of the above stated
correspondence, such a system is equivalent to a regular plane waveguide
filled by a gas with a spatially periodic permittivity tensor. The
permittivity modulation period is therewith equal to the grating period. As
the chosen plane grating has an asymmetric profile, the
corresponding virtual volume grating of the permittivity turns out to be
noncentrosymmetrical and, thus, satisfies to the above imposed
requirements for displaying of the T-violating phenomena.

Let $\hat \varepsilon _{ij}\left( r,\omega \right) =1+\hat \chi _{ij}\left(
x,z\right) $ be the permittivity of the waveguide being considered with as a
periodic function with respect to $z$. As it has been stated above
, such a waveguide can be modeled by the waveguide with
the effective permittivity $\hat \varepsilon _{eff}\left( z,\omega \right)
=1+\hat \chi _{eff}\left( z\right) $ which is a periodic function of $z$ and
is independent on $x$. We can show it mathematically starting with the
Maxwell equations

\begin{equation}
curl curl \vec E\left( \vec r,\omega \right) -\frac{\omega
^2}{c^2}\hat \varepsilon \left( r,\omega \right) \vec E\left( \vec r,\omega
\right) =0
\end{equation}

Let an electromagnetic wave propagate in the plane regular waveguide. In this case, the motion along the $z$-axis is free and, thus,
it can be described by a plane wave $e^{ikz}$. It results in a set of
equations for determining of the stationary states of the waveguide:

\[
\frac{d^2\vec E_{n\alpha }\left( x\right) }{dx^2}+\kappa _n^2\vec E_{n\alpha
}\left( x\right) =0
\]

where $\kappa _n=\frac{2\pi n}d$ , $\alpha $ notes the polarization state of
the wave.

Then, the field $\vec E\left( \vec r,\omega \right) $ in the waveguide with
the diffraction grating can be presented by the expansion as follows

\[
\vec E\left( \vec r,\omega \right) =\sum_nc_{n\alpha }\left( z\right) \vec
E_{n\alpha }\left( x\right)
\]

Substitution of this expansion into (1.50) yields

\begin{equation}
\frac{\partial ^2}{\partial z^2}c_{n\alpha }\left( z\right) +\left( \frac{%
\omega ^2}{c^2}-\kappa _n^2\right) c_{n\alpha }\left( z\right) +\frac{\omega
^2}{c^2}\sum_{\tau \alpha }\hat \chi _{\tau \alpha \alpha ^{^{\prime
}}}^{nn^{^{\prime }}}e^{-i\tau z}c_{a^{^{\prime }}}\left( z\right) =0
\end{equation}

where $\hat \chi _{\tau \alpha \alpha ^{^{\prime }}}^{nn^{^{\prime }}}=\int
\left\langle E_{n\alpha }\left( x\right) \left| \chi _{_\tau} \left( x\right)
\right| E_{n^{^{\prime }}\alpha ^{^{\prime }}}(x)\right\rangle dx$, $\chi _{_\tau}
=\frac 1a\int_0^a\chi \left( z\right) e^{i\tau z}dz$, $a$- is the period of
the plane grating.

Assuming the inequality $\frac{\chi ^{nn^{^{\prime }}}\frac{\omega ^2}{c^2}}{%
\kappa _n^{^{\prime }2}-\kappa _n^2}\ll 1$ , $\chi _{nn}\ll 1$ to be valid, we can
restrict ourselves to a single mode approximation, which allows us to
rewrite (1.50) as

\begin{equation}
\frac{\partial ^2}{\partial z^2}c_{n\alpha }\left( z\right) +\left( \frac{%
\omega ^2}{c^2}-\kappa _n^2\right) c_{n\alpha }\left( z\right) +\frac{\omega ^2%
}{c^2}\hat \chi _{\alpha \alpha ^{^{\prime }}}\left( z\right)
c_{\alpha^{^{\prime }}}\left( z\right) =0
\end{equation}

where $\hat \chi _{\alpha \alpha ^{^{\prime }}}\left( z\right) =\sum_{\tau}
\hat \chi _{\tau \alpha \alpha ^{^{\prime }}}^{nn^{^{\prime
}}}e^{-i\tau z}$.

Equations (1.52) are identical with those (see (1.10)) describing the electric
field in a volume diffraction grating. So, the goal formulated above has
been achieved: we have obtained a volume diffraction grating.

Let us remind (see (1.18)) that the diffraction grating manifests the optical
anisotropy (birefringens if the matter of grating possesses optical isotropy).
The anisotropy makes difficult to observe the T-violating photon
polarization plane rotation. Luckely, however, if the matter of the
waveguide or diffraction grating is a dielectric one can change the optical
anisotropy characteristics and even decrease the anysotropy sharply
by the electric field due to the Kerr or Pockels effect.

Now, let us estimate the effect of the T-violation. To do that we must
determine, in accordance with formula (1.47) for $\vartheta ^T$, the
T-violating susceptibility $\chi _{s1,2}^T$, which, in view of (1.13,1.3,1.4),
is proportional to the T-violating atomic polarizability $\beta _s^T$. The
estimate carried out by (Baryshevsky (1993,1994), Khriplovich (1991)) gives
$\beta _s^T\sim 10^{-3}\div10^{-4}\beta _s^P$, where $\beta _s^P$ is the
P-violating T-invariant scalar polarizability. The polarizability $\beta
_s^P $ was studied both theoretically and experimentally (Khriplovich
(1991)). Particularly, the theory gives $\beta _s^P\cong 10^{-30}cm^3$ for
atoms analogous to Bi, Tl, Pb. It yields the estimate $\cong
10^{-33}\div10^{-34}cm^3 $ for the T-violating atomic polarizability. The
polarizabelity $\beta_s^P $  causes the
P-violating rotation of the polarization plane by the angle $\vartheta^P =k%
 Re \chi _s^P\left( 0\right) L\cong 10^{-7}$ rad/cm$\times $L for
the gas density $\rho =10^{16} \div 10^{17}$. As a result, in our case the parameter $\varphi
=k\chi _s^T\left( \tau \right) L $ turns out to be $\varphi \cong
10^{-10}\div10^{-11}$ rad/cm$\times $L and can be even less by the factor h/d,
where h is the corrugation amplitude of the diffraction grating while d is
the distance between waveguide's mirrors. Assuming this factor to be ~$\sim
10^{-1}$, we shall find $\varphi \cong 10^{-11}\div10^{-12}$ rad/cm$\times $L.
Thus, the final estimate of the T-violating rotation angle $\vartheta ^T$ is

\begin{equation}
\vartheta ^T\cong 10^{-11} \div 10^{-12} {k_0}^2
{\chi _s^2 \left( \tau \right)} L^3
\end{equation}

In real situation the susceptibility of a grating $\chi _s\left( \tau
\right) $ may exceed the unity. However, our analysis has been performed
under the assumption $\chi _s\ll 1$. If, for example, we take $\chi
_s=10^{-1}$ , $k_0=10^4$ then $\vartheta ^T\simeq 10^{-6}\div10^{-7} L^3$
and, consequently, for L= 1 cm we will have the rotation angle
$\vartheta ^T \simeq 10^{-6} \div 10^{-7} rad$.

As it is seen, we have obtained the T-violating rotation angle $\vartheta ^T$
of the same order of $\vartheta ^P$. It makes possible experimental
observation of the phenomenon of the T-violating polarization plane rotation.

It should be noted that the manufacturing of diffraction gratings for the
range being more longwave than the visible light one may be simpler. That is
why we would like to attract attention to the possibility of studying of the
T-violating polarization plane rotation in the vicinity of frequencies of
atomic (molecular) hiperfine transitions; for example, for Ce (the
transition wavelength is $\lambda =3.26$ cm) and Tl ( $\lambda =1.42$ cm).

\subsection *{Conclusion}

Thus, we have shown that the phenomenon of the T-violating polarization
plane rotation appears while the photon is scattered by a volume diffraction
grating. The phenomenon grows sharply in the vicinity of the resonance
transmission condition. An experimental scheme based on a waveguide,
containing a diffraction grating and gas, has been proposed
which enables real experiments on observation of the T-violating
polarization plane rotation to be performed. The rotation angle has been
shown to be $\vartheta ^T=10^{-6} \div 10^{-7} L^3$, where L is
the waveguide length (thickness of the equivalent volume diffracting
grating).
Let there is the electric field in the region of the gas. As a result, the
scattering amplitude contains the T-violating term
$f_E^T (0)=i\beta_E^T \vec E \left[ \vec e^{~*} \times \vec e \right]$.
The phenomenon of the T-violating photon polarization plane rotation about
$\vec E$ appeares due to this term. The effect grows sharply in the volume
diffraction grating too, but the diffraction grating can be
centrisymmetrical. More over the effect can be observed in usual resonator
or ring laser because the sign of the photon polarization plane rotation
about $\vec E$ do not depend from the direction of the photon momentum
$\vec k$ unlike the P-violating, T-even polarization plane rotation effect
which is proportional to
$i \beta_v^P \vec n \left[ \vec e^{~*} \times \vec e \right]$.

\section {Neutron spin rotation and spin dichroism in media. }
\setcounter {equation}{0}

Neutron spin rotation and spin dichroism in media caused by parity (P)
violation and possible time (T) noninvariance under neutron interaction with
nuclei are being actively explored recently (Alfimenkov (1983),
Sushkov,Flambaum (1982), Stodolsky (1982), Koster et al (1991),
Baryshevsky (1983), Bowman et al (1990), Franke et al (1991)). The mentioned
effects are described by the refractive index of neutrons in media

\begin{equation}
\hat N=1+\frac{2\pi \rho }{k^2}\hat f(0),
\end{equation}
where $k$ is the neutron wave number, $\rho $ is the scatters density, $\hat
f(0)$ is the coherent elastic forward scattering amplitude of neutrons on a
nucleus.

M.Forte shows (Forte (1983), Forte, Zayen (1989)) that the T-noninvariant
effect of spin rotation arises under neutron diffraction by crystals with
nonpolarized nuclei. This effect is determined by coherent elastic scattering
amplitude at a non-zero angle $%
\hat f\left( \vec k^{^{\prime }},\vec k\right) \sim \vec \sigma \left( \vec
k^{^{\prime }}-\vec k\right) $ , where $\vec \sigma $ is the Pauli matrix
describing neutron spin, $\vec k^{^{\prime }}$ is the wave vector of
scattered neutron.

In (Forte (1983), Forte, Zayen (1989), Fedorov, Voronin, Lapin (1992))
the possibility of experimental measurements of this spin rotation
effect was considered. One of the main difficulties of the diffraction
methods suggested in
(Forte (1983), Forte, Zayen (1989), Fedorov, Voronin, Lapin (1992))
is that even for thermal neutrons the beams with
very low angular divergence rad and very high degree of monochromaticity $%
\frac{\Delta k}k\sim \frac{\Delta v}{v_B}\sim 10^{-5}\div 10^{-6}$ are
needed, where $v_B$ is the angle of Bragg diffraction. Only very small part $%
\Delta \Phi $ of a neutron flow $\Phi $ can be used to carry out the
experiment: $\Delta \Phi \sim \Phi \frac{\Delta k}k\Delta v$. The situation
is still worse for the range of epithermal neutrons in which $\Delta v\sim
10^{-7}$ rad and even lower.

In (Baryshevsky (1995))
we have shown that there are new effects: the effects of
T-noninvariant spin rotation and spin dichroism in the nonpolarized crystals
even in case when diffraction conditions are not fulfilled. As a
consequence, not only flow $\Delta \Phi $ increases, but the possibility of
investigation of in $\hat f\left( \vec k^{^{\prime }},\vec k\right) $ the
range of epithermal neutrons arises.

So, let the neutrons move through a crystal. According to (Baryshevsky
(1995), Baryshevsky (1976), Baryshevsky, Cherepitsa (1985)) the
Schrodinger equation describing propagation of a coherent neutron wave $\Psi
$ in a crystal with polarized nuclei in the impulse representation has the
view

\begin{equation}
\left( k^2-k_0^2\right) \Psi \left( \vec k\right) +\sum_{\vec \tau }\frac{2m%
}{\hbar ^2}\hat U_{eff}\left( \vec \tau \right) \Psi \left( \vec k-2\pi \vec
\tau \right) =0,
\end{equation}
where $k_{0}$ is the neutron wave number in vacuum, $k$ is the
neutron wave number in a crystal. The Fourie transform of neutron effective
potential energy in a crystal is

\begin{equation}
\hat U_{eff}\left( \vec \tau \right) =-\frac{2\pi \hbar ^2}{mV_0}\hat
F\left( \vec \tau \right) ; \hat F\left( \vec \tau \right)
=\sum_j\hat f_j\left( \vec \tau \right) e^{-w_j\left( \vec \tau \right)
}e^{-i2\pi \vec \tau \vec R_j},
\end{equation}
$\hat F\left( \vec \tau \right) $ is the amplitude of neutron coherent
scattering on crystal unit cell in the direction $\vec k^{^{\prime }}=\vec
k+2\pi \vec \tau $ , where $2\pi \vec \tau $ is the vector of crystal
reciprocal lattice, $\hat f_j\left( \vec \tau \right) $ is the amplitude of
coherent elastic scattering $\hat f_j\left( \vec k^{^{\prime }},\vec
k\right) $ on the nuclei of j-type in the direction $\vec k^{^{\prime
}}=\vec k+2\pi \vec \tau $ , $\vec R_j$ is the coordinate of the j nucleus
in the unit cell, $e^{-w_j\left( \vec \tau \right) }$ is the Debay-Waller
factor.

The system of homogeneous equations (2.2) permits to determine the dependence
of $k$ on $k_0$, i.e. to determine the neutron refractive index in a crystal.

It is well-known that the area $\Delta \upsilon $ of neutron incidence angle
on a crystal in which strong diffraction is observed (when the diffracted
wave amplitude is comparable with the amplitude of incident wave) is very
small $\Delta \upsilon \sim 10^{-5}\div 10^{-6}$ rad even for thermal
neutrons.

Outside this narrow angle area the diffracted wave amplitude is small and
the system of equations (2.2) can be analyzed according to the perturbation
theory. As a result, we have

\begin{equation}
\hat N=\frac k{k_0}\simeq 1+\frac 12\hat g\left( 0\right) +\frac{\hat
g\left( -\vec \tau \right) \hat g\left( \vec \tau \right) }{2\alpha _B},
\end{equation}

where

\[
\hat g\left( \vec \tau \right) =-\frac{2m}{\hbar ^2k_0^2}\hat U_{eff}\left(
\vec \tau \right) ,\alpha _B=\frac{2\pi \vec \tau \left( 2\pi \vec \tau
+2\vec k_0\right) }{k_0^2}
\]

As we see, the correction to the neutron refractive index in crystals
contains the scattering amplitude at the non-zero angle $\hat f\left( \vec
\tau \right) $.

According to (Baryshevsky (1995)) the expression for
$\hat g\left( \vec \tau \right) $ can be written as a sum

\begin{equation}
\hat g\left( \vec \tau \right) =\hat g_s\left( \vec \tau \right) +\hat
g_{so}\left( \vec \tau \right) +\hat g_w\left( \vec \tau \right) ,
\end{equation}
where $\hat g_s\left( \vec \tau \right) $ is proportional to the scattering
amplitude on the nucleus $\hat f_s\left( \vec \tau \right) $ due to strong
interactions minus spin-orbital scattering interactions; $\hat g_w\left(
\vec \tau \right) $ is proportional to the P and T-violating amplitude; $%
\hat g_{so}=\hat g_{son}+\hat g_{shw}$ , $\hat g_{son}$ is proportional to
the amplitude of spin-orbital neutron scattering on the nucleus due to
nuclear forces $\hat f_{son}\left( \vec \tau \right) =f_{son}\vec \sigma
\left[ \vec k\times 2\pi \vec \tau \right] $ , $\hat g_{shw}$ is
proportional to the amplitude of spin orbital neutron scattering on the
nucleus due to neutron magnetic momentum interaction with the nucleus
electric field (the so-called Schwinger scattering):

\begin{equation}
\hat g_{shw}=i\frac{2m}{V_0\hbar ^2k_0^2}\frac{\mu \hbar }{mc}\sum_j\Phi
_j\left( \vec \tau \right) e^{-w_j\left( \vec \tau \right) }\vec \sigma
\left[ \vec k\times 2\pi \vec \tau \right] e^{-i2\pi \vec \tau \vec \tau _j},
\end{equation}
$\mu $ is the magnetic neutron momentum, $\Phi _j\left( \vec \tau \right) $
is the Fourie component of electrostatic potential induced by nucleus $j$.

In case of the target with nonpolarized nuclei $\hat g_s\left( \vec \tau
\right) =g_s\left( \vec \tau \right) $ does not depend on neutron spin.
Besides, for slow neutrons $g_{so}\ll g_s$ . It allows to write the
contribution to caused by diffraction in the following way

\begin{equation}
\delta \hat N=\frac{\hat g\left( -\vec \tau \right) \hat g\left( \vec \tau
\right) }{2\alpha _B}=\frac{g_s\left( -\vec \tau \right) g_s\left( \vec \tau
\right) }{2\alpha _B}+\frac 1{2\alpha _B}g_s\left( -\vec \tau \right) \left[
\hat g_{so}\left( \vec \tau \right) +\hat g_w\left( \vec \tau \right)
\right] +
\end{equation}

\[
+\frac 1{2\alpha _B}g_s\left( \vec \tau \right) \left[ \hat g_{so}\left(
-\vec \tau \right) +\hat g_w\left( -\vec \tau \right) \right] .
\]

Let the target be nonpolarized. In this case T-noninvariant part of $\hat
g\left( \vec \tau \right) $ is (Baryshevsky (1995))
\begin{equation}
\hat g_w^T\left( \vec \tau \right) =\frac{4\pi }{k^2V_0}\sum_j\hat
f_{jw}^T\left( \vec \tau \right) e^{-w_j\left( \vec \tau \right) }e^{-i2\pi
\vec \tau \vec R_j},
\end{equation}

\[
\hat f_{jw}^T\left( \vec \tau \right) =C_j^{^{\prime }}\left( \vec \tau
\right) \vec \sigma \frac{\vec \tau }\tau , C_j^{^{\prime }}\left(
\vec \tau \right) =ReC_j^{^{\prime }}+ImC_j^{^{\prime }}.
\]

From (2.7,2.8) we see that the T-noninvariant contribution to the refractive
index in a nonpolarized crystals may occur only in a noncentrosymmetric
crystals.

We see also that we have the two effects: the difference of the absorption
coefficient of the neutrons with spin parallel and antiparallel to the
vector $\vec \tau $ (spin dichroism) and the spin rotation around of the
direction of $\vec \tau $.

Difference $\ae _{\uparrow \uparrow }-\ae _{\downarrow \uparrow }=k(Im\delta
N_{\uparrow \uparrow }-Im\delta N_{\downarrow \uparrow })$

Spin rotation angle $\upsilon =kRe\left( \delta N_{\uparrow \uparrow
}-\delta N_{\downarrow \uparrow }\right) L$

Difference of the neutrons number $J_{\uparrow \uparrow }$ with spin
parallel to $\vec \tau $ and the neutrons number $J_{\downarrow \uparrow }$
with spin antiparallel to $\vec \tau $ which passed through the crystal is

\begin{equation}
\frac{J_{\uparrow \uparrow }-J_{\downarrow \uparrow }}{J_{\uparrow \uparrow
}-J_{\downarrow \uparrow }}=kIm\left( \delta N_{\uparrow \uparrow }-\delta
N_{\downarrow \uparrow }\right) L
\end{equation}
where $L$ is the length of the neutron way in the crystal.

From (2.7) it follows that at the constant $\alpha _B$ {the angle }$\upsilon
$ {decreases proportionally to }$k^{-3}$ {and absorption }$\sim k^{-1}$ {%
with the neutron energy increasing. So, at first glance the transfer to the
epithermal range only makes worse the situation. However, it should be taken
into account that in the range of resonances the amplitudes }$f_w\left( \vec
\tau \right) ${sharply increase due to amplification mechanisms, being
analogous to the well-known ones for the scattering amplitudes
(Forte (1983), Baryshevsky (1995)).
According to (Alfimenkov (1983)) there are two amplification mechanisms:
dynamical and kinematic . They lead to the increase of the scattering
amplitude near resonance in
}$\sqrt{\frac{\Delta E}D}\cdot \sqrt{\frac{\Gamma _{ns}}{\Gamma
_{np}}}\simeq 10^5\div 10^6${, where }$\Delta E$ {is the averaged distance
between the single particle nuclei states, }$D$ {is the averaged distance
between the levels of the compound nucleus,} $\Gamma _{ns(p)}$ {are the
neutron widths of }$s(p)$ {the neutron state in the incident wave. As a
result, for the neutrons with the energies }$1-10ev$ {we can expect the
effect increase in }$10^3\div 10^2$ {times and as a consequence the decrease
in limitation for the detection of T- noninvariant effect. It is very
important that neutron part }$\Delta \Phi $ {even increases as the magnitude
}$\frac{\Delta k}k\sim \frac{\Delta \upsilon _B}{\upsilon _B}$ {and for }$
\Delta \upsilon _B=10^{-4}$ , $\upsilon _B=\frac{2\pi \tau }k\sim 10^{-1}${
we have ten time increase of }$\Delta \Phi ${. At the same time
investigation methods of spin rotation in the field of strong diffraction
for the neutrons with such an energy are practically unrealized, since
angular range of strong diffraction becomes }$\Delta \upsilon _B\sim 10^{-7}$
{rad. Two reasons simultaneously spoil the effect: angular spread of crystal
mosaic structure (}$\Delta \epsilon _m\geq 10^{-6}$ {rad) and sharp
suppression of flux }$\Delta \Phi $ {in such a small angular and energetic
interval }$\Delta \upsilon _B\frac{\Delta k}k\sim \frac{\Delta \upsilon _B^2%
}{\upsilon _B}=10^{-13}${.}

{\ It is very important to say that the difference of absorption coefficient
changes with the neutron energy increasing more slowly (}$\sim k^{-1}$) {%
than }$\upsilon $ {and sharply increases in the range of resonances also. As
a result the study of the effect of spin dichroism may give the essential
advantage in epithermal energy range. }

\newpage
\section*{References}

\hskip 0.6 cm
To part 1.

{ Christenson J.H., Cronin J.W., Fitch V.L. and Turlay R.  Phys. Rev.
Lett. $\bf{13}$ (1964) 1138}

{ V.G.Baryshevsky  Phys. Lett A 177 (1993) 38

{ V.G. Baryshevsky and D.V.Baryshevsky   Journ of Phys. B: At. Mol.
Opt. Phys. $\bf{27}$ (1994) 4421}

{ A.M.Barkov  and N.S.Zolotariov   Lett. J. Exp. Theor. Phys. $\bf{118}$
(1978) 379 (in Russian)}

{ M.A.Bouchiat  and L.Pottier  Science $\bf{234}$ (1986) 1203}

{ V.V.Fedorov , V.V.Voronin , and E.G.Lapin Journ. of Phys. G:
Nucl.Part. Phys. G $\bf{18}$ (1992) 1133}

{ M.J.Forte  Journ. of Phys. G.: Nucl. Part. Phys. G $\bf{9}$ (1983) 745}

{ J.D.Jackson   Classical electrodinamics (John Wiley \& Sons, INC,
New York - London) (1962)}

{ I.V.Khriplovich  Party Nonconservation in Atomic Phenomena 1991 (London:
Gordon and Breach)

{ S.K.Lamoreaux   Nucl. Instrum. Methods A $\bf{284}$ (1989) 43}

{ S.A.Maksimenko, G.Ya.Slepyan  Electromagnetics $\bf{17}$ (1997) 147}

{ Shih-Lin Chang  Multiple Diffraction of X-rays in Crystals 1984
(Springer-Verlag Berlin Heidelberg New York Tokyo)}

{ T.Tamir  Guided-Wave Optoelectronics  1988 (Springer Verlag New York)}
\vskip 1cm

To part 2.

{ V.P.Alfimenkov, Nuclear Physics A398 (1983) 93.}

{ O.Sushkov, V.Flambaum, Usp.Fiz.Nauk 25 (1982) 1.}

{ L.Stodolsky, Phys.Lett. B197 {1982) 213. } }

{ J.E.Koster et al, Phys.Lett B267 (1991) 23. }

{ V.G.Baryshevsky, Sov.J.Nucl.Phys. 38 (1983) 699. }

{ J.Bowman et al, Phys.Rev.Lett.65 (1990) 1192. }

{ C.Franke et al, Phys. Rev.Lett 67 (1991) 564. }

{ M.Forte, J. of Phys.G 9 (1983) 745. }

{ M.Forte, C.Zayen, Nucl. Instr. Met. A284 (1989) 147. }

{ V.Fedorov, V.Voronin, E.Lapin, J. of Phys.  Journ. of Phys. G:
 Nucl.Part. Phys. G $\bf{18}$, (1992), 1133}

{ V.G.Baryshevsky, Physics of Atomic Nuclei 58 (1995) 1471 }

{ V.G.Baryshevsky, J Exp.Theor Phys. 70 (1976) 430 }

{ V.G.Baryshevsky, S.V.Cherepitsa, Phys.Stat.Sol (b) 128 (1985) 379 }

{ L.I.Lapidus, Rev.Nod.Phys. 39 (1967) 689 }

\end{document}